\documentclass[twocolumn]{aastex63}
\usepackage{graphics,epsf}
\usepackage{amsmath}                
\usepackage{amsfonts}               
\usepackage{amssymb}                
\usepackage{epsfig}                 
\usepackage{appendix}
\usepackage{graphicx}
\usepackage{float}
\usepackage{color}
\usepackage{multirow}
\usepackage{colortbl}
\usepackage[para,online,flushleft]{threeparttable}
\usepackage{url}

\hypersetup{citecolor=blue, 
            linkcolor=red, 
            menucolor=blue, 
            urlcolor=blue}  
            
\usepackage{hyperref}

\newcommand{\cm}{{~\rm cm}}

\newcommand{\km}{{~\rm km}}
\newcommand{\s}{{~\rm s}}

\newcommand{\g}{{~\rm g}}

\newcommand{\erg}{{~\rm erg}}
\newcommand{\yr}{{~\rm yr}}

\newcommand{\days}{{~\rm days}}

\begin{document}

\title{Feeding post core collapse supernova explosion jets with an inflated main sequence companion}


\author{Ofek Hober}
\affiliation{Department of Physics, Technion, Haifa, 3200003, Israel; ealeal44@technion.ac.il; soker@physics.technion.ac.il}

\author{Ealeal Bear}
\affiliation{Department of Physics, Technion, Haifa, 3200003, Israel; ealeal44@technion.ac.il; soker@physics.technion.ac.il}

\author{Noam Soker}
\affiliation{Department of Physics, Technion, Haifa, 3200003, Israel; ealeal44@technion.ac.il; soker@physics.technion.ac.il}

\begin{abstract}
We simulate the response of a main sequence star to the explosion of a stripped-envelope (type Ib or Ic) core collapse supernova (CCSN) when the main sequence star orbits the core at a distance of $10 R_\odot$ or $20 R_\odot$ at explosion. We use the stellar evolution code \textsc{mesa} to follow the response of main sequence stars of masses $3 M_\odot$ and $7M_\odot$ to energy deposition and mass removal. The collision of the CCSN ejecta with the main sequence star deposits energy and inflate the main sequence star. If the binary system stays bound after the CCSN explosion the inflated main sequence star might engulf the newly born neutron star (NS). We assume that the NS accretes mass through an accretion disk and launches jets. The jets remove mass from the inflated main sequence star and collide with the CCSN ejecta. Although this scenario is rare, it adds up to other rare scenarios to further support the notion that many stripped envelope CCSNe are powered by late jets. The late jets can power these CCSNe-I for a long time and might power bumps in their lightcurve. The jets might also shape the inner ejecta to a bipolar morphology. Our results further support suggestions that there are several ways to feed a NS (or a black hole) to launch the late jets in superluminous supernovae.  
\end{abstract}


\section{INTROCUTION}
\label{sec:intro}

The total radiated energy of super luminous supernovae (SLSNe) can be $\ga 10^{50} \erg$ (e.g., \citealt{GalYam2012, Wangetal2016, Arcavietal2016, Sorokinaetal2016, Chenetal2017b, LiuModjaz2017, DeCiaetal2018, Marguttietal2018, Linetal2020, Lunnanetal2020, Moriyaetal2020, Chenetal2021}; for a recent review of observations see \citealt{Nicholl2021}). There are several possible energy sources to explain SLSNe (e.g., \citealt{Wangetal2019RAA}). One energy source model to account for the extra energy is a magnetar, i.e, a rapidly spinning-magnetic neutron star (NS) remnant of the core collapse supernova (CCSN; e.g., \citealt{Greineretal2015, Metzgeretal2015, Kangasetal2017, Kasenetal2016, Chenetal2017a, Mazzalietal2017, Nicholletal2017a, Yuetal2017, Villaretal2018, ChatzopoulosTuminello2019, Cartieretal2021, Nicholl2021, Gomezetal2022}; for a partial list of other energy sources see, e.g., \citealt{Urvachevetal2021} and \citealt{Lietal2020}).

However, despite being popular in recent years literature, the magnetar model has some limitations. Firstly, even if magnetars can account for late lightcurves of SLSNe, in most cases jets must power the explosion itself \citep{Soker2016Mag, Soker2017Mag2, SokerGilkis2017, Soker2022LSNe} as neutrino driven explosion mechanism cannot account for the required explosion energy of $E_{\rm ex} \ga 2 \times 10^{51} \erg$ (e.g., \citealt{Fryeretal2012, SukhboldWoosley2016}).
 Secondly, the magnetar cannot explain bumps in the lightcurve of some CCSNe. 
\cite{Hosseinzadehetal2021} noted recently that magnetar are not expected to lead to bumps in the lightcurve of many type Ib and type Ic CCSNe. These hydrogen-poor CCSNe are called stripped-envelope supernovae (SESNe) or CCSNe-I. On the other hand, other recent studies suggest that late (days to months after explosion) jets are very efficient in channelling kinetic energy to radiation and in forming bumps in the lightcurve of CCSNe (e.g., \citealt{KaplanSoker2020, AkashiSoker2021, Soker2022bump}). 
\cite{GofmanSoker2019} argue that the pure-magnetar powering model of the extraordinary Type II supernova iPTF14hls (\citealt{Arcavietal2017, Sollermanetal2019}) that \cite{Woosley2018} suggested, better works if jets add to the powering of the lightcurve. 
\cite{KaplanSoker2020Jetshaped} suggest that jets better explain the abrupt decline in the lightcurve of the hydrogen-poor luminous SN~2018don than a magnetar model that \cite{Lunnanetal2020} apply to that CCSN. \cite{Soker2022LSNe} argue that the inclusion of jets better explains luminous supernovae (LSNe) than the magnetar-based modelling that \cite{Gomezetal2022} apply to LSNe. 

There are many studies that mention the possible roles of jets in CCSN explosions (e.g., \citealt{Wheeleretal2002, Milisavljevic2013, Lopezetal2014, BrombergTchekhovskoy2016, Inserraetal2016, Chenetal2017a, Mauerhanetal2017, Reilletal2017,Barnesetal2018}). The vast majority of these and other studies consider jets to play a role only in rare cases of CCSNe. Some give a larger role to jets, although not in most CCSNe, e.g., \cite{Sobacchietal2017} who speculate that relativistic jets power all SESNe. We, on the other hand, take it that \textit{jets explode the majority, or even all, CCSNe} (e.g., \citealt{PapishSoker2011, GilkisSoker2015, Bearetal2017, BearSoker2017, GrichenerSoker2017, Soker2017RAA, ShishkinSoker2021, ShishkinSoker2022, Soker2022Boosting}).
Jets can account for other properties of some CCSNe.  \cite{Gilkisetal2016} argued that jets can explain the high luminosity of some CCSNe, including SESNe, that the commonly cited neutrino-driven explosion model cannot explain, e.g., as the CCSNe that \cite{Prenticeetal2021} and \cite{Sollermanetal2022} studied recently. 
 
To form a magnetar, i.e., a rapidly rotating magnetic NS, the pre-collapse core must have a large amount of angular momentum. We expect this to be the case only if a companion spins-up the pre-collapse core. For that a companion should be close enough to the core to make tidal interaction significance, or to merge with the core. In this study, we consider a main sequence companion at an orbital separation of several solar radii (low mass main sequence stars) to several tens of solar radii (high mass main sequence stars) at the time of the explosion.  Such pre-collapse binary systems might be the progenitors of low and high mass X-ray binaries (e.g., \citealt{ZuoLi2014, Wiktorowiczetal2017, vandenHeuvel2019}). 
The collision of the ejecta with the companion during the explosion deposits energy to the envelope of the companion (e.g., \citealt{LiuZWetal2015}) and causes its expansion (e.g., \citealt{Hiraietal2018}) to the degree that it can transfer mass to the newly born NS or black hole (e.g., \citealt{Rimoldietal2016, Ogataetal2021}).   
 
This study continues our exploration of the role of late jets in CCSNe. When a rapidly rotating core collapses to form a NS or a black hole it also forms an accretion disk that might last for a long time around the newly born NS or black hole (e.g., \citealt{Gilkis2018}). This disk is very likely to launch two opposite jets in one or more jet-launching episodes  (e.g., \citealt{Nishimura2015, Chenetal2017a}). In this study we consider another mass source for the accretion disk that launches the jets. We consider CCSNe that explode with a close main sequence companion. The explosion ejecta inflates the main sequence star to the degree that it transfers mass to an accretion disk around the NS, or around a black hole if the NS collapses to a black hole. This accretion disk in turn launches the jets. We use the stellar evolutionary code \textsc{mesa} (section \ref{sec:Numerical}) to inflate main sequence companions and to remove mass to mimic the interaction of the inflated star with the NS remnant of the CCSN (section \ref{sec:Inflated}). The newly born NS might end on an eccentric orbit inside the inflated envelope (section \ref{sec:Orbit}). In section \ref{sec:jets} we discuss the implications of the mass transfer from the envelope to the NS to the launching of late jets that form bumps in the lightcurves of some SESNe. We summarise in section \ref{sec:summary}.

\section{Numerical Scheme}
\label{sec:Numerical}
\subsection{The MESA subroutine}
\label{subsec:Numerical}
We follow the evolution of $M_{2,0}=3M_\odot$ and $7M_\odot$ zero age main sequence stellar models as we deposit energy and remove mass as a consequence of CCSN explosion. For that we use the stellar evolution code Modules for Experiments in Stellar Astrophysics (\textsc{mesa}; \citealt{Paxtonetal2011, Paxtonetal2013, Paxtonetal2015, Paxtonetal2018, Paxtonetal2019}), version 10398 in its single star mode.
We divide our simulation to two stages as follows.
In stage A we follow the \textit{inlist} (the input parameters for \textsc{mesa}) that \cite{Ogataetal2021} used and published in Zenodo \footnote{ \url{https://zenodo.org/record/4624586\#.Yi2rn3pBxPY}}.  
The metallicity is $z=0.02$ as in \cite{Ogataetal2021}.
We follow the evolution of the star from the pre main sequence evolution up to the zero age main sequence, or very close to it, when the star reaches a minimum radius. The minimum radii are
$R_{2,0}(3M_\odot) = 2.0R_\odot$ and $R_{2,0} (7 M_\odot)= 3.2R_\odot$.  

In stage B we follow the \textit{inlist} that \cite{Ogataetal2021} used in depositing energy after the explosion. We take the explosion energy as $E_{\rm expl}=10^{51} \erg$ and deposit the energy into the main sequence star during a time period of $\Delta t_{\rm dep} = 0.95\yr$ as we describe in section \ref{subsec:Energy_subroutine}. The separation of the main sequence and the core at explosion is either $a_i=10 R_\odot$ or $a_i=20 R_\odot$. 
 
We differ from \cite{Ogataetal2021} in that we remove mass from the envelope of the main sequence star. We remove mass at a constant rate starting from the time we start to deposit the energy. The mass removal rates are either  
$\dot M_2 = - 0.01M_\odot \yr^{-1}$ or $\dot M_2=- 0.05M_\odot \yr^{-1}$  (see discussion in section \ref{sec:Inflated}).  

\subsection{The energy subroutine}
\label{subsec:Energy_subroutine}
In the injection of energy into the envelope we follow \cite{Ogataetal2021} who fit the results of the numerical simulations of \cite{Hiraietal2018}. Their prescription for energy deposition to the envelope is as follows. 
 
The energy deposition per unit mass in the outer envelope mass of $m_{\rm h}$ is constant, and it declines with decreasing radius linearly with the mass measured from the surface inward, $m$.  
The mass $m_{\rm h}$ is half the mass of the ejecta that hits the main sequence companion 
\begin{equation}
m_{\rm h} = \frac{\tilde{\Omega} M_{\rm ej}} {2} ,
\label{eq:mh}
\end{equation}
where $M_{\rm ej}$ is the mass of the ejecta which we here take to be $M_{\rm ej}=2 M_\odot$,
\begin{equation}
\tilde{\Omega} = \frac{\Omega}{4 \pi} =
\frac{1}{2} \left[ 1- \sqrt{1-\left( \frac{R_{2,0}}{a_i} \right)^2}  \right]
\label{eq:tildeOmega}
\end{equation}
is the fractional solid angle of the companion looking from the primary, $R_{2,0}$ is the radius of the companion at the moment of the explosion and $a_i$ is the distance between the collapsing core and the companion at the explosion moment. 
The fraction of the kinetic energy of the ejecta that hits the companion that ends in the companion envelope is $p \simeq 0.08-0.1$, such that the total energy that the ejecta deposits to the companion envelope is 
\begin{equation}
E_{\rm heat}=p \tilde \Omega E_{\rm expl}, 
\label{eq:Eheat}
\end{equation}
where $E_{\rm expl}$ is the total explosion energy (kinetic energy of the ejecta).  
\cite{Ogataetal2021} take the energy that the ejecta-companion interaction deposits per unit mass to be  
\begin{equation}
\Delta \epsilon(m)= \frac {E_{\rm heat}}{m_{\rm h}} 
\frac {{\rm min}(1,m_{\rm h}/m)}{1+\ln (M_2/{m_{\rm h}})} .
\label{eq:DeltaEm}
\end{equation}
Note again that the mass $m$ is measured from the surface inward.

The ejecta deposits most of the energy within less than a day. However, due to numerical limitations \cite{Ogataetal2021} add the energy to the companion in a much longer timescale of close to a year. We use the same procedure and take for the energy deposition period $\Delta t _{\rm dep}=0.95 \yr$. 

A comment is in place here on the injection time period. The CCSN ejecta hit the main sequence star and deposit energy to the main sequence star within hours. However, the dynamical time of the star becomes much larger. As we will see in section \ref{sec:Inflated}, in our simulations the final stellar radius without mass removal and just after we end energy injection is $\ga 300 R_\odot$. The dynamical time of the inflated star becomes $\simeq 0.3 \yr$. We cannot assume an expansion under hydrostatic equilibrium (as the numerical code assumes) within few days. It should last for few months rather. Because the NS starts close to the main sequence ($a_i=10R_\odot$ or $a_i=20R_\odot$) very early on during the expansion the NS starts to accrete mass and as a result of that, under our assumption, to launch jets. The jets then deposit energy to the envelope. Therefore, it might be that the energy that we should deposit to the envelope is larger even than what we deposit here. However, without a three-dimensional hydrodynamical simulations we cannot estimate this energy to better accuracy. 
The above considerations justify our scheme of starting to inject energy and remove mass at $t=0$ and to continue for about one year. 

\section{Envelope inflation and mass loss}
\label{sec:Inflated}

We describe the response of two main sequence stellar models to the energy deposition and mass removal from the main sequence stars as we described in section \ref{sec:Numerical}. We assume that the main sequence companion is at an initial orbital separation of either $a_i=10 R_\odot$ or $a_i=20 R_\odot$ from the core at the moment of explosion, and that the CCSN explosion energy is $E_{\rm exl}=10^{51} \erg$. 

We consider cases where after explosion the binary system of the newly born NS and the main sequence star stays bound. The post-explosion semi-major axis might increase to tens of solar radii and more, and therefore the orbital period might increase to be about a week to few months, depending on the natal kick velocity of the NS, the mass that the explosion ejects, and the mass of the main sequence star. 

The collision of the CCSN ejecta with the main sequence star inflates the main sequence star such that it engulfs the NS (\citealt{HiraiYamada2015}; see below). The removal of mass by the NS from the inflated main sequence star might take several orbits, or even tens of orbits. In any case, for numerical reasons we cannot remove mass too rapidly. We examine the influence of the mass removal rate from the main sequence star by examining two rates of mass removal, $\dot M_2=-0.01 M_\odot \yr^{-1}$ and of $\dot M_2=-0.05 M_\odot \yr^{-1}$. 
These mass removal rates come from the following consideration. We assume that jets that the NS launches remove the envelope mass above the initial orbital separation, $M_{\rm env} (r>a_i)$,  in a timescale of several weeks to several months, up to about a year (sections \ref{sec:Orbit} and \ref{sec:jets}). As we show below, this mass amounts to  $M_{\rm env}(r>10R_\odot) \simeq 0.02-0.04 M_\odot$. Therefore, the appropriate mass loss rate is $\approx 0.02-0.1 M_\odot \yr^{-1}$. However, the NS might acquire a highly eccentric orbit (section \ref{sec:Orbit}) and spends most of the time at distance of $r \gg 10 R_\odot$, where the envelope mass is much lower and mass removal time longer. We therefore consider the two mass loss rates above. 
The numerical code \textsc{mesa} removes mass from the outer layer of the star and the mass carries with it its energy.
We also note that we do not change the mass of the main sequence star as a result of the collision of the CCSN ejecta with it \citep{Ogataetal2021}.

When the inflated envelope radius decreases below the periastron distance the NS ceases to accrete mass from the envelope at a high rate. We, nonetheless, follow the evolution of the main sequence star under mass removal to later times. Note that the energy deposition  time period is from $t=0$ to $t=\Delta t_{\rm dep}=0.95 \yr$.  

\begin{figure}
\includegraphics[trim= 3.5cm 9.4cm 0.0cm 9.4cm, scale=0.63]{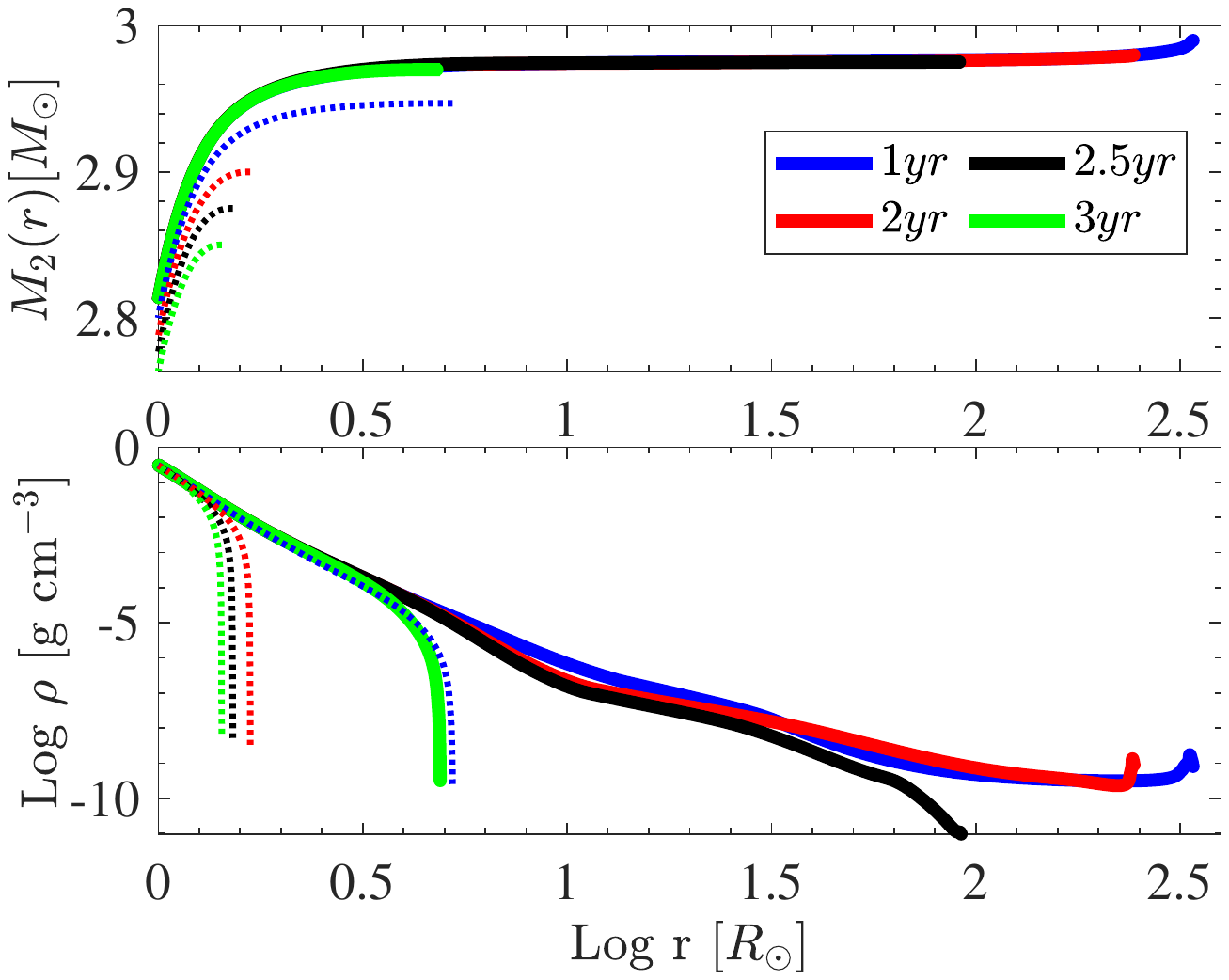}
\caption{Mass versus radius (upper panel) and density versus radius (lower panel) for the $M_{2,0}=3M_\odot$ stellar model and for an initial separation of the NS and the main sequence star of $a_i = 10 R_\odot$. The energy deposition lasts for $\Delta t_{\rm dep} = 0.95\yr$ starting at $t=0$, while the mass removal continues for the entire simulation. The solid lines represent a mass removal rate of  $\dot{M}_2 = -0.01M_\odot \yr^{-1}$ from the inflated main sequence star and the dotted lines represent the case with $\dot{M}_2 = -0.05M_\odot \yr^{-1} $. The blue, red, black and green lines are at times of $1\yr$, $2\yr$, $2.5\yr$ and $3\yr$ after the SN explosion. Note that the star achieves the largest inflated radius at the end of the energy deposition, just before the first time we present here  (see Fig. \ref{Fig:R_vs_T_3yr}). }
 \label{Fig:3Mo_ao_10Ro}
 \end{figure}

\begin{figure}
\includegraphics[trim= 3.5cm 9.4cm 0.0cm 9.4cm, scale=0.63]{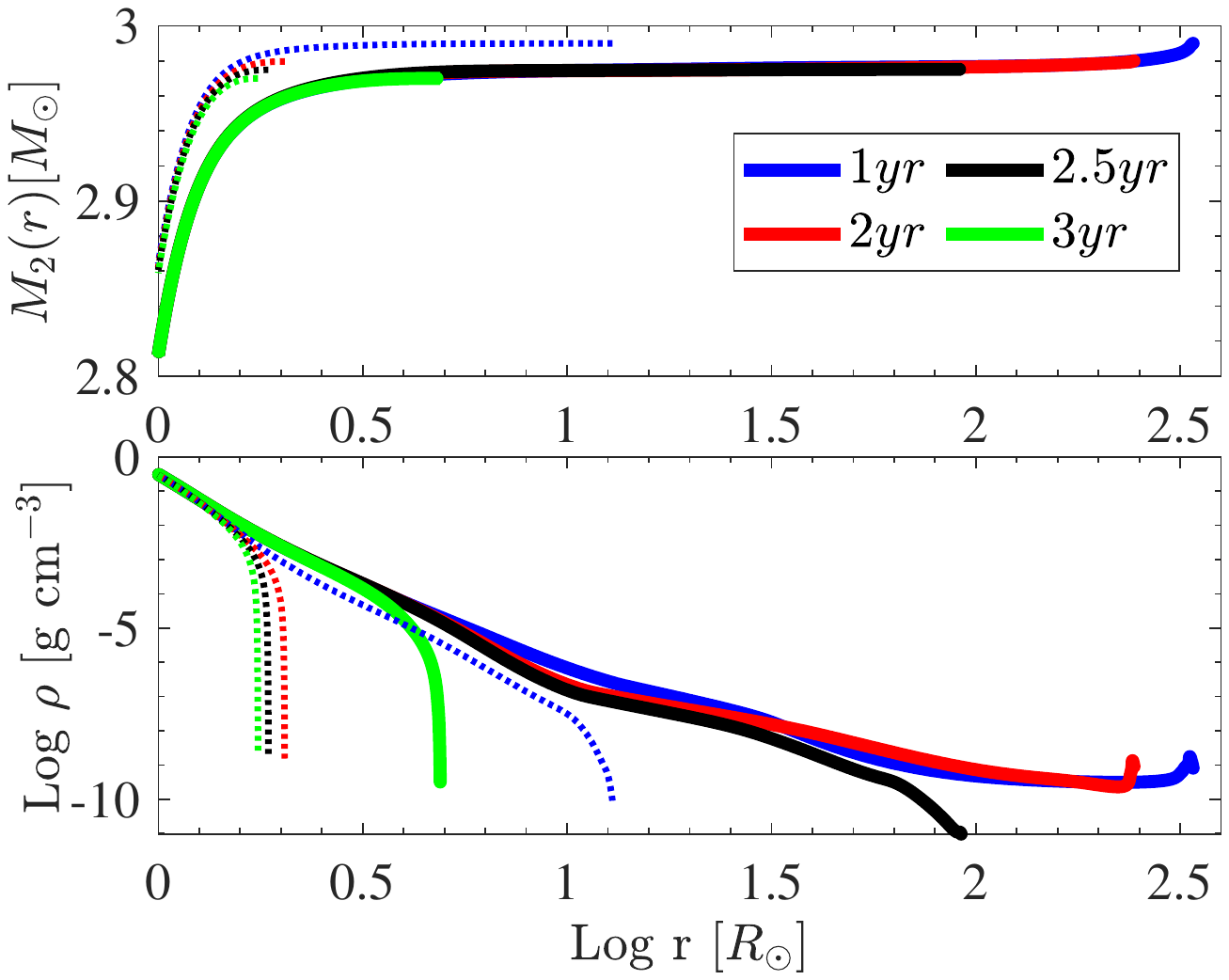}
\caption{Similar to Fig. \ref{Fig:3Mo_ao_10Ro}. The solid lines are identical to those in Fig. \ref{Fig:3Mo_ao_10Ro}, i.e., initial separation of $a_i=10 R_\odot$, while the dotted lines represent an initial orbital separation of $a_i = 20 R_\odot $.
In both cases $M_{2,0}=3M_\odot$ and $\dot M_2 = -0.01M_\odot \yr^{-1}$. 
 }
 \label{Fig:3Mo_Mdot_0_01}
 \end{figure}

\begin{figure}
\includegraphics[trim= 3.5cm 9.4cm 0.0cm 9.4cm, scale=0.63]{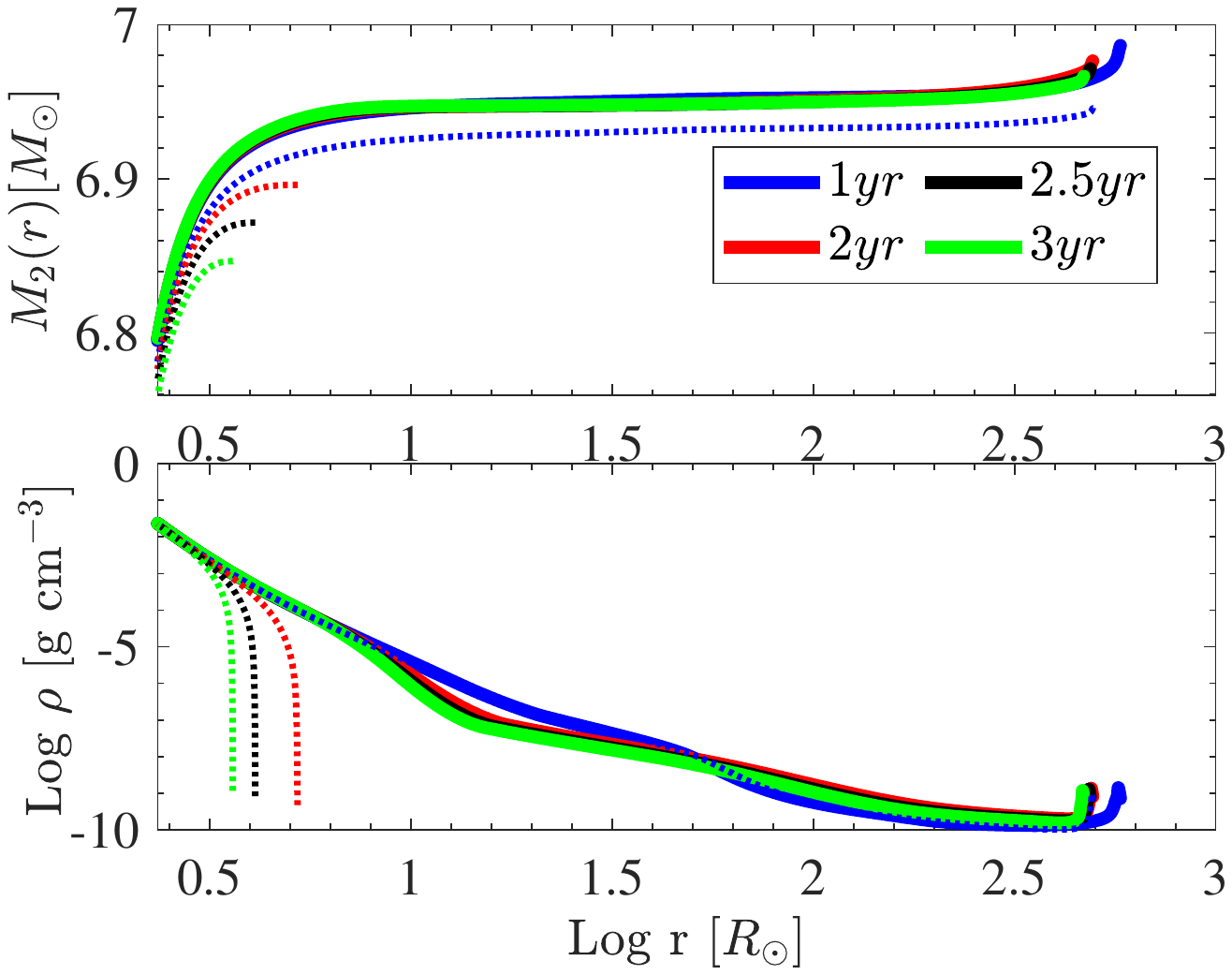}
\caption{Similar to Fig. \ref{Fig:3Mo_ao_10Ro}, but for the cases with $M_{2,0} =7M_\odot$. As in Fig. \ref{Fig:3Mo_ao_10Ro}, $a_i = 10R_\odot$, and the solid lines are for $\dot{M}_2 = -0.01M_\odot \yr^{-1}$ and the dotted lines are for $\dot{M}_2 = -0.05M_\odot \yr^{-1} $. }
 \label{Fig:7Mo_a0_10Ro}
 \end{figure}

In Fig. \ref{Fig:3Mo_ao_10Ro} we present the mass $M_2(r)$ inside a radius $r$ as function of $r$ (upper panel) and the density profile $\rho (r)$ at four times for the case with $(M_{2,0},a_i)=(3 M_\odot, 10 R_\odot)$. We present the results for the two mass removal rates. We present only the region of $r>1 R_\odot$.  In Fig. \ref{Fig:3Mo_Mdot_0_01} we compare two cases with different initial orbital separations. The increase in $a_i$ reduces the amount of energy that we deposit to the envelope according to equations (\ref{eq:tildeOmega}) and (\ref{eq:Eheat}). 
In Fig. \ref{Fig:7Mo_a0_10Ro} we present the behavior of the $(M_{2,0},a_i)=(7 M_\odot, 10 R_\odot)$ for the two mass removal rates. 

\begin{figure}
\includegraphics[trim= 3.5cm 9.4cm 0.0cm 9.4cm, scale=0.63]{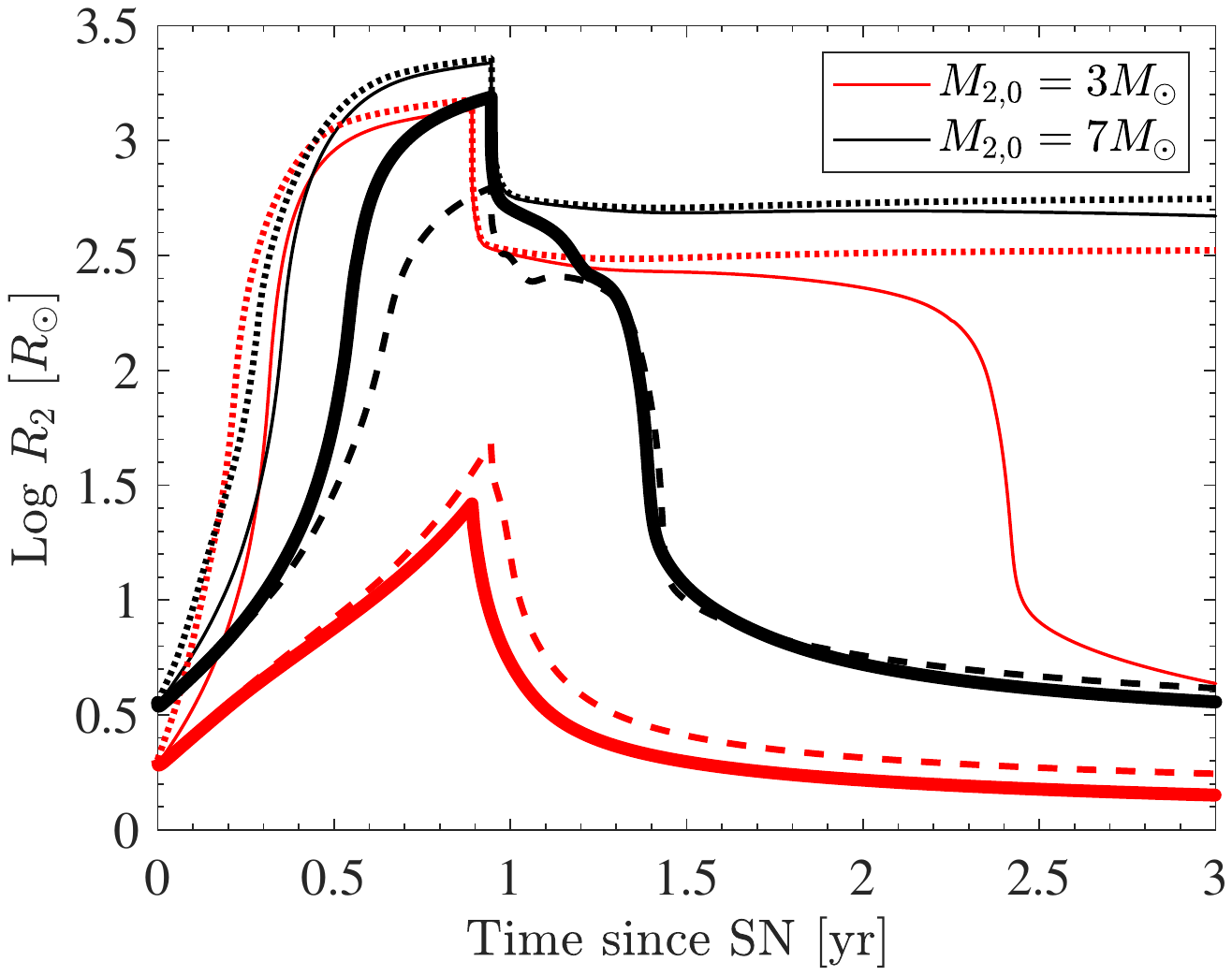}
\caption{Radius versus time since the SN explosion at $t=0$. We deposit energy from $t=0$ to $t=0.95 \yr$. The dotted lines include no mass removal. In the other cases we remove mass at a constant rate from $t=0$ to end of the plots. The thick lines represent a mass loss of rate of $\dot{M} = -0.05M_\odot \yr^{-1}$ and the thin lines represent a mass loss rate of $\dot{M} = -0.01M_\odot \yr^{-1}$. 
Solid lines are for an initial separation between the exploding core and the main sequence star of $a_i=10 R_\odot$ and the dashed lines are for $a_i = 20 R_\odot $. The red lines are for   $M_{2,0}=3M_\odot$ and the black lines are for $M_{2,0}=7M_\odot$.}
 \label{Fig:R_vs_T_3yr}
 \end{figure}

In Fig. \ref{Fig:R_vs_T_3yr} we present the evolution of the stellar radius with time for the different cases we simulate.
Although we remove mass at a constant rate for the entire duration of the simulations, once the radius of the inflated main sequence star gets smaller than the periastron distance the high mass loss rate substantially decreases. Because we assume a pre-explosion circular orbit, in case of no natal kick velocity to the NS the periastron distance would be $a_i$. Otherwise it can be smaller than $a_i$ (unless the kick velocity is exactly along the pre-explosion velocity of the NS).   
  
Some of the cases we simulate, specifically, the cases  $(M_{2,0},a_i,\dot M_2)=(3 M_\odot, 10 R_\odot,- 0.05 M_\odot \yr^{-1})$ and $(M_{2,0},a_i,\dot M_2)=(3 M_\odot, 20 R_\odot,- 0.01 M_\odot \yr^{-1})$, are inconsistent. The reason is that the secondary radius (the two lower red lines in Fig. \ref{Fig:R_vs_T_3yr}) stays larger than the initial separation for only a short time. Therefore, we do not expect much removal of mass from the inflated main sequence envelope. Namely, we overestimate the mass removal rate in these two cases. The correct value of mass removal is less extreme, and the envelope expands to larger radii and for a longer time. 

According to Figs. \ref{Fig:3Mo_ao_10Ro} - \ref{Fig:7Mo_a0_10Ro} the typical density in the inflated main sequence star envelope is $\rho \approx 10^{-7} \g \cm^{-3}$ at $r \approx 10 R_\odot$ and down to $\rho \approx 10^{-9} \g \cm^{-3}$ in the outer inflated zones. 

In post-explosion bound binaries the orbit is eccentric. Therefore, as the main sequence star contracts the NS gets in and out of the MS envelope. Therefore, although the phase of largely inflated envelope might last for about two years in some of the cases that we simulate, the accretion phases combined might be shorter. 
 Later we scale the relevant jet-launching episode with a time period of $\tau_{\rm j} = 0.05 \yr$. 

The main conclusion from the evolution of the main sequence star under energy deposition and mass removal is that in many systems that stay bound after explosion the NS orbits inside the inflated envelope of the main sequence star where the density is $\rho \approx 10^{-9} - 10^{-7} \g \cm^{-3}$, and for several months accretes mass. In section \ref{sec:jets} we turn to discuss the implication of this situation. 
  
\section{Post-explosion orbits}
\label{sec:Orbit}

\subsection{Orbital properties}
\label{subsec:OrbitalProperties}

For our proposed scenario we require that the post-explosion binary system of the newly born NS and the inflated main sequence star stays bound. This will be the output in a large fraction of cases, although not in all. We consider few examples. In all cases we take for the ejecta mass $M_{\rm ej}=2 M_\odot$ as we mentioned in section \ref{subsec:Energy_subroutine}, and the NS mass $M_{\rm NS}=1.4 M_\odot$. 
Typical natal kick velocities in binaries are $v_{\rm k} \simeq 100 \km \s^{-1}$ (e.g., \citealt{Fortinetal2022}). Lower values are possible as well.

Consider two cases with $M_2=3 M_\odot$ and $a_i=10 R_\odot$, one with no natal kick and one with.  
For these two cases we find the post-explosion eccentricity, apastron distance, and orbital period of the binary system to be $e_{\rm p}=0.45$, $r_{\rm max,p}=27 R_\odot$  and $P_{\rm p}=4.3~{\rm days}$, respectively, for the case without natal kick, and $e_p=0.90$, $r_{\rm max,p}=191 R_\odot$  and $P_{\rm p}=56~{\rm days}$ for a case with a natal kick of $v_{\rm k} =50 \km \s^{-1}$ along the direction of the NS orbital motion at the moment of explosion. For a natal kick velocity of $v_{\rm k} >60 \km \s^{-1}$ along the initial orbit of the NS this system becomes unbound. This limited kick velocity increases if the kick is opposite to the initial orbital motion. 

For the case $M_2=7 M_\odot$ and $a_i=10 R_\odot$ we find that the system becomes unbound for $v_{\rm k} >120 \km \s^{-1}$. For $v_{\rm k} = 50 \km \s^{-1}$ and $v_{\rm k} = 100 \km \s^{-1}$ along the orbital motion of the NS and for $M_2=7 M_\odot$ and $a_i=10 R_\odot$, we find $e_{\rm p}=0.53$ $r_{\rm max,p}=33 R_\odot$  and $P_{\rm p}=3.9~{\rm days}$   and $e_{\rm p}=0.86$ $r_{\rm max,p}=129 R_\odot$  and $P_{\rm p}=23~{\rm days}$, respectively.  

For the case $M_2=7 M_\odot$ and $a_i=20 R_\odot$ we find that the system becomes unbound for $v_{\rm k} >85 \km \s^{-1}$. For $M_2=7 M_\odot$ and $a_i=20 R_\odot$ we find the post-explosion orbital parameters to be $e_{\rm p}=0.24$ $r_{\rm max,p}=33 R_\odot$ $P_{\rm p}=5.4~{\rm days}$ for a case without a kick velocity, and $e_{\rm p}=0.66$ $r_{\rm max,p}=99 R_\odot$ and $P_{\rm p}=18~{\rm days}$, for $v_{\rm k} = 50 \km \s^{-1}$ along the orbital motion of the NS.

 These several examples demonstrate that in many cases the systems might stay bound. These examples also serve for our scaling in the next section.  

\subsection{Orbital evolution}
\label{subsec:OrbitalEvolution}
From the inflated envelope radius (Fig. \ref{Fig:R_vs_T_3yr}) and the properties of the orbit that we derive for several examples in section \ref{subsec:OrbitalProperties} we conclude that the NS spirals-in inside the envelope for a time period that might last from about few weeks to several months. This amounts to about several to several tens orbits. During that time the NS suffers gravitational friction from the inflated envelope around it and tidally interacts with the envelope inner to its orbit. 

We estimate that these effects are small for the following reasons. 
First, the total envelope mass outside the initial orbit of the NS, where it is likely to orbit most of the time, is very low, $M_{\rm env}(r>10R_\odot) \simeq 0.02-0.04 M_\odot < 0.03 M_{\rm NS}$. Therefore, the envelope cannot remove much orbital energy nor orbital angular momentum. Note  also that the total interaction time is several to several tens times the orbital time period. Tidal interaction needs a longer time to significantly change the orbit in such a large ratio of the NS mass to the outer envelope mass. 

Moreover, we are interested in the power of the jets. Any strong interaction will remove envelope mass, hence reducing the jets' power. Therefore, the relevant jet interaction takes place before a significant envelope mass is removed from the inflated parts.  

Despite the small effects we discussed above, they operate to shorten the duration of the common envelope evolution (CEE) of the NS inside the inflated envelope. The same goes to the effect of the jets that remove envelope mass. During part of the time the jets will be deflected and decelerated inside the inflated envelope. For all these effects we will scale the relevant time that the NS launches jets that expand to interact with the CCSN ejecta with a time scale of only $\tau_{\rm j}=0.05 \yr$. Namely, a time scale much shorter than the timescale of the CEE if we do not consider the above effects (which is several weeks to about a one-two years).

\section{Implication to jet launching}
\label{sec:jets}

This section combines the new results of section \ref{sec:Inflated} regarding the inflated envelope properties as a response to energy deposition and mass removal and the estimates of the orbital properties from section \ref{sec:Orbit} to present new estimates of the jets' power and timescale (section \ref{subsec:Properties}). We then consider the new results of section \ref{subsec:Properties} in the context of previous numerical hydrodynamical simulations of jet-ejecta interaction (section \ref{subsec:Effects}). These numerical studies \citep{AkashiSoker2020, AkashiSoker2021, AkashiSoker2022}, however, only assumed some jets properties. Here, we actually calculate the jets' properties. Our new results regarding the jets' properties motivate more detail hydrodynamical simulations of jets-ejecta interaction. Moreover, our study motivate the much more complicated 3D simulations of CEE with a NS that launches jets. The community does not have yet the full capability to perform a full CEE with energetic jets. 

\subsection{The jets' power and timescale}
\label{subsec:Properties}

\cite{Hiraietal2018} show that the ejecta-companion interaction has a very small influence on the orbit. The NS natal kick might have a large influence on the orbit. Because in the present study the ejecta mass is smaller than the combined mass of the NS and the secondary star, $M_{\rm ej} = 2M_\odot < M_{\rm NS} + M_2$, i.e., less than half the initial mass of the binary system, we assume that in many cases the system stays bound  (section \ref{subsec:OrbitalProperties}). 
Therefore, even if the orbit is highly eccentric, at periastron the distance of the NS from the main sequence star is several to about twenty solar radii for the parameters that we simulate here. 
\cite{Hiraietal2018} further mention the possibility that the main sequence transfers mass to the NS and even engulfs the NS after the explosion \citep{HiraiYamada2015}. We here consider the possibility that the NS accretes mass through an accretion disk and launches jets. 

\cite{Ogataetal2021} conducted a thorough study of the response of the companion to the ejecta with \textsc{MESA}. \cite{Ogataetal2021} discuss also several aspects of the presence of the companion, like its appearance years to hundreds of years post-explosion. They also mention that the accretion onto the NS can unbind the inflated envelope of the main sequence companion. They, however, did not discuss the role of jets. In previous sections we followed their prescription for ejecta-companion interaction, but added mass loss that mimic the accretion of mass by the NS and the ejection of envelope mass by the jets that the NS launches. We here discuss the possible properties of these jets. 

The flow structure we deal with after the inflation of the main sequence stellar envelope is of a CEE where a NS orbits inside a giant low-density envelope. The NS accretes mass through an accretion disk and launches jets. The jets remove envelope mass in a negative feedback cycle (for a review of this jet feedback mechanism see \citealt{Soker2016Rev}). 

We can only crudely estimate the Bondi--Hoyle--Lyttleton (BHL) mass accretion rate by the NS. The reason is that the orbit of the NS inside the inflated envelope of the main sequence star depends on the mass lost in the explosion and the natal kick of the NS (section \ref{subsec:OrbitalProperties}).  In particular, the orbits might be highly eccentric. This in turn implies that the inflated-envelope density from which the NS accretes mass changes along the orbit, as does the relative envelope-NS velocity. Therefore, the mass accretion rate, and as a result of that the jets' power, might vary along the NS orbit. Near apastron the NS might even be outside the inflated-envelope, i.e., the NS star launches jets as it accretes mass from the envelope in a Roche lobe overflow process (unless the apastron to envelope radius ratio is too large).

We write a scaled equation where we assume that the relative envelope-NS velocity is the relative NS orbital velocity in the envelope as if the  orbit is circular (although it is not) and with a radius of $a > a_i$. To the accuracy of our treatment we can neglect the sound speed with respect to the orbital motion. 
Scaling by a typical density from Figs. \ref{Fig:3Mo_ao_10Ro} - \ref{Fig:7Mo_a0_10Ro} the BHL mass accretion rate is 
\begin{eqnarray}
\begin{aligned}
& \dot M_{\rm acc,BHL} \simeq  0.03     
\left( \frac{M_{\rm NS}}{1.4 M_\odot} \right)^2 
\left( \frac{M_{\rm NS}+M_2}{4.4 M_\odot} \right)^{-3/2} 
\\ & \times 
\left( \frac{a}{50 R_\odot} \right)^{3/2} 
\left( \frac{\rho}{10^{-8} \g \cm^{-1}} \right) 
M_\odot \yr^{-1} ,
\end{aligned}
\label{eq:Macc}
\end{eqnarray}
where $M_{\rm NS}$ is the NS mass and $M_2$ the main sequence mass. 

We examine the possibility that the NS accretes mass and launches jets. We adopt claims by earlier studies (e.g.,  \citealt{ArmitageLivio2000, Papishetal2015, SokerGilkis2018}) that because of the very small radius of the NS the specific angular momentum of the accreted mass, as a result of accretion from the non-homogeneous envelope, is sufficiently large to form an accretion disk. The question then is the accretion rate. First we note that the accretion process is highly non-spherical and the medium is optically thick. Therefore, we cannot apply directly the Eddington luminosity limit. 
In addition, neutrino cooling allows a very high mass accretion rate when the mass accretion rate is $\dot M_{\rm acc} \ga 10^{-3} M_\odot \yr^{-1}$ \citep{HouckChevalier1991, Chevalier1993, Chevalier2012}.  

Numerical simulations show that the mass accretion rate is somewhat smaller than the BHL accretion rate, by a factor of $\xi \simeq 0.1-0.5$ (e.g., \citealt{Livioetal1986, RickerTaam2008, MacLeodRamirezRuiz2015a,  MacLeodRamirezRuiz2015b, Chamandyetal2018, Kashietal2022}). In addition, the jets reduces the mass accretion rate by a factor of $\chi_{\rm j} \simeq 0.1-0.2$ \citep{Gricheneretal2021, Hilleletal2022}. The jets themselves carry a fraction $\eta$ of the accretion energy. 
  
One possible outcome is a feedback cycle. Namely, there is a phase of accretion at a high rate when neutrino cooling does take place that launches jets. The jets then turn off the accretion, and so the jets cease to exist. This allows the high mass accretion rate to resume itself, starting the next cycle. Overall, the power of the jets is 
\begin{eqnarray}
\begin{aligned}
& \dot E_{\rm 2j}  = \eta \xi \chi_{\rm j} \dot M_{\rm acc,BHL} \frac {G M_{\rm NS}} {R_{\rm NS}} 
=  1.8 \times 10^{42} 
\left( \frac{\eta}{0.2} \right) 
\\ & \times 
\left( \frac{\xi}{0.3} \right) 
\left( \frac{\chi_{\rm j}}{0.1} \right) 
\left( \frac{\dot M_{\rm acc,BHL}}{0.03M_\odot \yr^{-1}} \right) 
\erg \s^{-1} ,
\end{aligned}
\label{eq:E2j}
\end{eqnarray}
where $R_{\rm NS} =12 \km$ is the NS radius and we substituted $M_{\rm NS}=1.4 M_\odot$ above.  
Because the mass accretion rate, $\xi \chi_{\rm j} \dot M_{\rm acc,BHL}$, is just a factor of few larger than the rate for which neutrino cooling is efficient, we took in equation (\ref{eq:E2j}) the jets to carry a relatively large fraction, $\eta \simeq 0.2$,  of the accretion energy.

Another effect is that the NS eccentric orbits brings it to inner regions inside the main sequence inflated envelope, where the density can become as high as $\rho \simeq 10^{-7} \g \cm^{-1}$. 

The mass in the inflated envelope, i.e., above the orbital radius at explosion $a_i$, is $M_{\rm env}(r>a_i) \simeq 0.02-0.04 M_\odot$. However, because of the jet feedback mechanism the NS accretes only a small fraction of this mass before the entire inflated envelope is removed, mainly by the jets. The binding energy of the inflated envelope of the $M_2=3M_\odot$ secondary star that is at $a_i=10 R_\odot$ (density profiles in Fig. \ref{Fig:3Mo_ao_10Ro}) is $E_{\rm bind} \approx 10^{46} \erg$ (depending on the inner radius that we take for the ejected envelope). 
 Despite that the energy that the jets carry (see below) is much larger than the binding energy of the envelope, we expect that in some cases, although not all, the process will continue for weeks and longer. These cases will be when the post-explosion apastron distance is large such that the NS spends most of the time at $r \ga 50 R_\odot$. The reasons are as follows. 

In the outer regions the jets that the NS launches remove the outer zones of the inflated envelope. Near apastron the system might perform the grazing envelope evolution (GEE). In the GEE the jets remove the outer zones of the envelope and accelerate them to acquire large energies, much above the escape energy (e.g., \citealt{Shiberetal2017}). While \cite{Shiberetal2017} simulated the GEE with jets from a main sequence companion, here the jets are much faster, as in the common envelope jets supernova (CEJSN) impostor simulations of  \cite{Schreieretal2021}. As well, \cite{Schreieretal2021} simulate a NS on a highly eccentric orbit that enters the red supergiant envelope and gets out. We expect here also highly eccentric orbits.  \cite{Schreieretal2021} find that the jets that the NS launches in their CEJSN impostor eject envelope mass at velocities much above the escape speed. The unbound mass carry on average an energy which is tens time the binding energy.  
We expect the same here. 
As the envelope shrinks near apastron the NS might be outside the envelope. In that case the NS accretes via a Roche lobe overflow, and the jets can then expand more freely. Note that although the density in the NS orbit is similar in our case and in the simulations of \cite{Schreieretal2021}, the jets in our case need to penetrate through a lower envelope mass than in the simulations of \cite{Schreieretal2021}.

Since the NS launches the jets  perpendicular to the equatorial plane, the jets are not efficient at removing envelope mass in and near the equatorial plane. The jets might  break out from the inflated envelope and expand almost freely. This implies that the jets are very inefficient at removing the inflated envelope mass. In addition, the NS cannot remove the envelope within less than one orbit. For an orbital period of $P_{\rm orb}= 19.5 (a/50 R_\odot)^{3/2}[(M_{\rm NS} + M_2)/4.4 M_\odot]^{-1/2} \days$ and with the jets' power as in the scaling of equation (\ref{eq:E2j}) the total energy the jets carry in one orbit is $E_{\rm 2j}(P_{\rm orb}) \simeq 3 \times 10^{48} \erg$.
These values imply a jet feedback efficiency of $\beta_{\rm j} =E_{\rm bind}/E_{\rm 2j} \approx 0.1-1 \%$. 
However, our expectation is that the envelope mass decreases and the jets power decreases during a timescale longer than one orbit, about few orbital period. 

We note that our energy budget above is global. Namely, we take the binding energy of the inflated envelope above the initial orbital radius, i.e., the inflated envelope in the zone $r \ga a_i$. Although we scale the jets' power (equation \ref{eq:E2j}) at radius $r=50 R_\odot$, this is actually a global value as it represents the crude average jets' power. We deal, as stated above, mainly with cases where the post-explosion apastron distance is large. Therefore, in the relevant cases for our study the NS spends most of its time at large orbital separations of $r\ga 50 R_\odot$. However, the NS also dives to its initial orbital separation (section \ref{subsec:OrbitalProperties}) where it interacts with the envelope at those radii. Therefore, the NS jets need to remove also the envelope mass at those radii. 
We also note that although the density at $r=10 R_\odot$ is two orders of magnitude larger than the typical density at $r=50 R_\odot$, the accretion rate goes as $r^{-3/2}$. Therefore, the accretion rate at $r=10 R_\odot$ is only about one order of magnitude above than at $r=50 R_\odot$. In any case, the NS spends a short time at and near that radius because it is the periastron distance in a highly eccentric orbit.

From the discussion above and for the case of a secondary star of mass $M_2=3M_\odot$ that is at $a_i=10 R_\odot$ at explosion, we scale the jets' activity period by $\tau_{\rm j} \approx 0.05 \yr$, and the total energy that the jets carry by  
\begin{eqnarray}
\begin{aligned}
& E_{\rm 2j}  \approx  
2.8 \times 10^{48} 
\left( \frac{\eta}{0.2} \right) 
\left( \frac{\xi}{0.3} \right) 
\left( \frac{\chi_{\rm j}}{0.1} \right) 
\\ & \times 
\left( \frac{\dot M_{\rm acc,BHL}}{0.03M_\odot \yr^{-1}} \right) 
\left( \frac{\tau_{\rm j}}{0.05 \yr} \right)
\erg .
\end{aligned}
\label{eq:E2jTot}
\end{eqnarray}
We emphasize again that we expect the jet power to vary a lot along the orbit. 

We comment on the scaling of the jets' activity period timescale by $\tau_{\rm j} = 0.05 \yr \simeq 2.5~{\rm weeks}$.
We estimated above that without any interaction between the NS and the jets it launches with the inflated envelope the NS will spend several weeks to several months and up to about a year inside the inflated envelope. However, there are two effects here. First is the removal of envelope mass by the jets (which deposit much larger energy that the orbital energy by dynamical friction). This mass removal causes the inflated envelope to shrink. On the other hand the jets deposit energy to the inflated envelope. Although the unbound mass carry away most of this energy, some energy does stay in the still-bound envelope gas (e.g., \citealt{Schreieretal2021}). This extra energy will keep the envelope inflated for a longer time than what we find here with mass removal.
Without full 3D CEE hydrodynamical simulations that include a NS that launches jets, we cannot be more accurate. Therefore, the time scale might in some cases be only about one week, while in some other cases, probably when the post-explosion apastron is at a very large distance, the times scale might be several months. These uncertainties should be kept in mind while using the scaled equations we derive here.  

\subsection{The effects of the jets}
\label{subsec:Effects}

In section \ref{subsec:Properties} we conclude that for a time period of $\tau_{\rm j}$ of a few weeks the NS that orbits inside the inflated envelope of the main sequence star is likely to launch jets that carry a total energy of $E_{\rm 2j} \approx 10^{48} -10^{49} \erg$ (equation \ref{eq:E2jTot}). We now discuss the possible effects that such jets might have on the CCSN properties. 

\cite{AkashiSoker2020} conducted three-dimensional hydrodynamical simulations of jets that a NS companion to a type Ic or type Ib CCSN launches a few hours after explosion. The NS companion that is at about $5 R_\odot$ from the exploding core accretes mass from the ejecta, and launches jets that interact with the ejecta. The jets in these simulations were active from one hour after explosion to about three hours after explosion and the total jets' energies in these simulations were about $1 \%$ of the explosion energy of $10^{51} \erg$. \cite{AkashiSoker2020} found that the jets, which the NS companion launches off-center relative to the explosion, inflate one low-density hot bubble to one side of the ejecta. This hot bubble expands to shape ejecta up to expansion velocities of $\simeq 3500 \km \s^{-1}$. 

\cite{AkashiSoker2021} conducted three-dimensional hydrodynamical simulations of jets that the newly born NS launches weeks to months after explosion as it accretes fallback material from the ejecta. In those simulations the jets' energies were $1.6\times 10^{49} \erg$ and $4 \times 10^{50} \erg$, i.e., larger than the jets' energies we expect in the present scenario. 
  
\cite{AkashiSoker2022} simulated jets that the NS launches about half an hour after explosion as it accretes ejecta fallback material. They simulated cases with jets' energies from $4 \times 10^{45} \erg$ to $4 \times 10^{49} \erg$, and followed the jets as they propagate into the ejecta of a CCSN with an explosion energy of $4 \times 10^{51} \erg$ and ejecta mass of $5 M_\odot$. 

 The above three studies \citep{AkashiSoker2020, AkashiSoker2021, AkashiSoker2022} did not derive the jets properties, but rather assumed them. We here differ in estimating the jets properties (section \ref{subsec:Properties}). 

We scale the discussion of section 4 from \cite{AkashiSoker2022} to a CCSN of explosion energy of $E_{\rm SN}= 10^{51} \erg$. The CCSN ejecta expands in a homologous flow, namely the velocity increases linearly with the radius. We repeat their assumption that the jets influence the inner ejecta up to some velocity that we mark $v_{\rm ej,d}$. Like \cite{AkashiSoker2022} we take this velocity to be where the energy of the ejecta inner to that velocity is about ten times the energy of the jets, $E_{\rm ej} (<v_{\rm ej,d}) \simeq 10 E_{\rm 2j} $. Although the average magnitude of the effect of the jets on this inner ejecta is $10 \%$ by energy, along the polar directions (the directions of the two opposite jets) the influence of the jets on the flow is much larger. From their equations (5) and (6) we find that the jets have an average effect of $10 \%$ by energy on the inner region of the ejecta that contains a mass of 
\small
 \begin{eqnarray}
\begin{aligned}
\frac{ M_{\rm ej} (<v_{\rm ej,d}) }{M_{\rm ej}} 
& \simeq 0.16 
\left( \frac{E_{\rm SN}}{10^{51} \erg} \right)^{-1/2}
\\ & \times 
\left( \frac{E_{\rm 2j}}{2.5 \times 10^{48} \erg} \right)^{1/2}.
\end{aligned}
\label{eq:Mej(Vejd)}
\end{eqnarray}
\normalsize
 
The main differences of our new study from some earlier studies of post-explosion weak jets (e.g., \citealt{AkashiSoker2020, AkashiSoker2021, AkashiSoker2022, KaplanSoker2020, Soker2020}) are that in the present scenario (1) the feeding of the young NS is by an inflated main sequence companion, rather than feeding of an old or young NS by the ejecta material, and (2) that we actually estiamte teh jets' properties. We expect the jets that we propose in this study to have some similar effects to those that these earlier studies have simulated and discussed. 

The interaction of the jets with the ejecta efficiently converts kinetic energy to radiation (e.g., \citealt{KaplanSoker2020}). If the interaction takes place after the main peak of the CCSN lightcurve, then the jets might cause a `bump' in the CCSN lightcurve (e.g., \citealt{KaplanSoker2020}). In the present study the CCSNe are stripped-envelope CCSNe that ejects little mass the time to the main peak is short. Therefore, the jets-ejecta interaction that takes place weeks to months after explosion is likely to take place after the main peak in the lightcurve.
In the present study the typical ratio of the energy of one jet to the explosion energy is
\begin{equation}
\epsilon_E \equiv \frac{E_{\rm 2j}/2}{E_{\rm expl}} \approx \frac{1.4 \times 10^{48} \erg}{10^{51} \erg} = 1.4 \times 10^{-3}. 
\label{eq:Epsilon}
\end{equation}
\cite{KaplanSoker2020} built a toy model to fit the third bump (third peak) in the lightcurve of iPTF14hls with $\epsilon_E=7.9 \times 10^{-3}$. Their figure 6 shows that the bump maximum luminosity is about 10 times as luminous as the supernova lightcurve at that time. Namely, even for $\epsilon_E \simeq 10^{-3}$ they would have obtain a clear jet-powered bump in the lightcurve. 
\cite{Soker2022bump} uses the same toy model to fit the bump at t=110~days in the lightcurve of SN~2019zrk. 
\cite{Soker2022bump} finds that he could fit the bump with jets that have $\epsilon_E \simeq 6 \times 10^{-4}$. 
Therefore, we suggest that the jets that we propose here  which have $\epsilon_E \approx 10^{-3}$, might power a small bump, or up to few bumps, in the CCSN lightcurve weeks to months after explosion. 

The second effect is that the jets shape the inner regions of the ejecta to a bipolar structure. The fraction of the inner ejecta mass that the jets shape is crudely given by equation (\ref{eq:Mej(Vejd)}). Our study adds to the claim of \cite{AkashiSoker2022} that in a small fraction of CCSNe post-explosion weak jets shape the very inner zones of the ejecta to possess a bipolar structure. We differ from \cite{AkashiSoker2022} in the source of the jets and that they are launched at a much later time in the present study, but the shaping process is qualitatively similar. In both cases the outer parts might have the imprints of the jets that explode the CCSN, whether jittering jets or fixed-axis jets. 

As \cite{AkashiSoker2022} discussed, the shaping by jets influences also the distribution of elements in the inner regions of ejecta, and therefore has implications to the dispute on the processes behind the morphological features of clumps and filaments in some supernova remnants. While some studies attribute these morphological features to instabilities alone in the delayed neutrino mechanism (e.g., \citealt{Jankaetal2017, Larssonetal2021, Orlandoetal2021, Sandovaletal2021}), we attribute these morphological features to jets as well as instabilities, including the jets that explode the star in the jittering jets explosion mechanism and in some CCSNe also to post-explosion jets (e.g., \citealt{GrichenerSoker2017, Akashietal2018, BearSoker2018SN1987A, Soker2022}, and this study). 
We note though that the present study does not touch directly the dispute over the explosion mechanism of CCSNe (see section \ref{sec:intro}). 

\section{Summary}
\label{sec:summary}

Using the stellar evolutionary code \textsc{mesa} (section \ref{subsec:Numerical}) we simulated the response of main sequence stars to CCSN explosion at distances of $a_i=10 R_\odot$ and $a_i=20 R_\odot$. Such CCSNe are stripped-envelope CCSNe, namely SNe Ib or SNe Ic (CCSNe-I). The CCSN ejecta deposits energy into the main sequence star as it collides with it. We followed \cite{Ogataetal2021} in the procedure by which we deposit the energy into the main sequence star (section \ref{subsec:Energy_subroutine}). We departed from the simulations of \cite{Ogataetal2021} in also removing mass from the main sequence star. 

The deposition of energy to the main sequence star causes its expansion, and for the parameters we simulated here and if the binary system  stays bound after explosion the inflated envelope of the main sequence star engulfs the NS. We assumed that the NS accretes mass through an accretion disk and releases energy, mainly by launching jets. These jets remove mass from the inflated main sequence star envelope. We simulated two mass loss rates.
We presented the evolution of mass and density profiles in the evolving inflated envelope for several cases in Figs. \ref{Fig:3Mo_ao_10Ro} - \ref{Fig:7Mo_a0_10Ro}.  We also presented the evolution of the radius of the inflated envelope for several cases in Fig. \ref{Fig:R_vs_T_3yr}. 

From our simulations we learn that the NS might orbit inside the inflated main sequence envelope where densities are $\rho \simeq 10^{-9} - 10^{-7} \g \cm^{-3}$ for few weeks to several months (section \ref{sec:Orbit}). 
The accretion rate might be high enough (equation \ref{eq:Macc}) to allow for neutrino cooling, therefore allowing this high mass accretion rate as we study in section \ref{subsec:Properties}. The jets break out from the inflated main sequence envelope very rapidly, and deposit their energy, which we give in equation (\ref{eq:E2jTot}), into the CCSN ejecta. 

Sections \ref{sec:Inflated} and \ref{subsec:Properties} are our main new results.  
In section \ref{subsec:Effects} we combined our new results of jets' properties with previous 3D hydrodynamical numerical simulations of jets-ejecta interaction and with a toy model of bumps in the light curve. These simulations \citep{AkashiSoker2020, AkashiSoker2021, AkashiSoker2022} did not study the jets' properties, but rather assumed them.    
We followed the analysis by \cite{AkashiSoker2022} and estimated that the jets,  for which we derive the properties of in this study (section \ref{subsec:Properties}),  might shape the inner ejecta, $\approx 15 \%$ of the total ejecta mass (equation \ref{eq:Mej(Vejd)}), to a bipolar shape. 

In equation (\ref{eq:Epsilon}) we gave the ratio of the energy that one jet carries to the total explosion energy. Learning from previous studies that used a toy model to estimate the luminosity of jet-driven bumps in the lightcurve, we concluded that the jets we study here are energetic enough to induce a bump in the lightcurve at late times (after the peak). 

Our new results call for more detail 3D hydrodynamical simulations of the interaction of jets with CCSN ejecta, followed by the very demanding and challenging simulations of CEE of a NS that launches jets.  

On a broader scope, although the scenario we studied here of a newly born NS accreting from a main sequence is rare, when considering other rare cases we might expect a non-negligible number of CCSNe-I (type Ib and Ic CCSNe) to have late jets that power them and shape them. The other scenarios might be jets that a NS companion to the CCSN launches and accretion of fallback material (e.g., \citealt{Pellegrinoetal2022} for a recent discussion of some SNe Icn).

The jets might also power the lightcurve for a long time and in addition they might power a bump (small peak) to a few bumps in the lightcurve of these CCSNe, in particular after the main peak. Observations in recent years might support the notion that late jets power a large fraction of stripped envelope CCSNe (CCSNe-I), e.g., \cite{Soker2022LSNe}. Our study adds to the suggestions that there are several ways to feed a NS (or a black hole) to launch the late jets. 

\section*{Acknowledgements}
 
 We thank an anonymous referee for detailed comments that substantially improved the presentation of our results.  This research was supported by a grant from the Israel Science Foundation (769/20).

\section*{Data availability}
The data underlying this article will be shared on reasonable request to the corresponding author. 


\begin{thebibliography}

\bibitem[\protect\citeauthoryear{Akashi, Bear, \& Soker}{2018}]{Akashietal2018} Akashi M., Bear E., Soker N., 2018, MNRAS, 475, 4794. doi:10.1093/mnras/sty029

\bibitem[\protect\citeauthoryear{Akashi \& Soker}{2020}]{AkashiSoker2020} Akashi M., Soker N., 2020, ApJ, 901, 53. doi:10.3847/1538-4357/abad35

\bibitem[\protect\citeauthoryear{Akashi \& Soker}{2021}]{AkashiSoker2021} Akashi M., Soker N., 2021, MNRAS, 501, 4053. doi:10.1093/mnras/staa3897

\bibitem[Akashi \& Soker(2022)]{AkashiSoker2022} Akashi, M. \& Soker, N.\ 2022, \apj, 930, 59. doi:10.3847/1538-4357/ac6102


\bibitem[Arcavi et al.(2017)]{Arcavietal2017} Arcavi, I., Howell, D.~A., Kasen, D., et al.\ 2017, \nat, 551, 210

\bibitem[Arcavi et al.(2016)]{Arcavietal2016} Arcavi, I., Wolf, W.~M., Howell, D.~A., et al.\ 2016, \apj, 819, 35

\bibitem[Armitage \& Livio(2000)]{ArmitageLivio2000} Armitage, P.~J., \& Livio, M.\ 2000, \apj, 532, 540

\bibitem[Barnes et al.(2018)]{Barnesetal2018} Barnes, J., Duffell, P.~C., Liu, Y., et al.\ 2018, \apj, 860, 38. doi:10.3847/1538-4357/aabf84

\bibitem[Bear et al.(2017)]{Bearetal2017} Bear, E., Grichener, A., \& Soker, N.\ 2017, \mnras, 472, 1770

\bibitem[Bear \& Soker(2017)]{BearSoker2017} Bear, E., \& Soker, N.\ 2017, \mnras, 468, 140

\bibitem[Bear \& Soker(2018)]{BearSoker2018SN1987A} Bear, E. \& Soker, N.\ 2018, \mnras, 478, 682. doi:10.1093/mnras/sty1053


\bibitem[Bromberg \& Tchekhovskoy(2016)]{BrombergTchekhovskoy2016} Bromberg, O., \& Tchekhovskoy, A.\ 2016, \mnras, 456, 1739

\bibitem[\protect\citeauthoryear{Cartier et al.}{2021}]{Cartieretal2021} Cartier R., Hamuy M., Contreras C., Anderson J.~P., Phillips M.~M., Morrell N., Stritzinger M.~D., et al., 2021, arXiv, arXiv:2108.09828

\bibitem[Chamandy et al.(2018)]{Chamandyetal2018} Chamandy, L., Frank, A., Blackman, E.~G., et al.\ 2018, \mnras, 480, 1898

\bibitem[\protect\citeauthoryear{Chatzopoulos \& Tuminello}{2019}]{ChatzopoulosTuminello2019} Chatzopoulos E., Tuminello R., 2019, ApJ, 874, 68. doi:10.3847/1538-4357/ab0ae6

\bibitem[Chen et al.(2017a)]{Chenetal2017a} Chen, K.-J., Moriya, T.~J., Woosley, S., Sukhbold, T.,  Whalen, D.~J., Suwa, Y., \& Bromm, V.\ 2017a, \apj, 839, 85  

\bibitem[\protect\citeauthoryear{Chen et al.}{2021}]{Chenetal2021} Chen T.-W., Brennan S.~J., Wesson R., Fraser M., Schweyer T., Inserra C., Schulze S., et al., 2021, arXiv, arXiv:2109.07942

\bibitem[Chen et al.(2017b)]{Chenetal2017b} Chen, T.-W., Nicholl, M., Smartt, S.~J., et al.\ 2017b, \aap, 602, A9

\bibitem[Chevalier(1993)]{Chevalier1993} Chevalier, R.~A.\ 1993, \apjl, 411, L33

\bibitem[Chevalier(2012)]{Chevalier2012} Chevalier, R.~A.\ 2012, \apjl, 752, L2

\bibitem[De Cia et al.(2018)]{DeCiaetal2018} De Cia, A., Gal-Yam, A., Rubin, A., et al.\ 2018, \apj, 860, 100. doi:10.3847/1538-4357/aab9b6

\bibitem[\protect\citeauthoryear{Fortin et al.}{2022}]{Fortinetal2022}  Fortin F., Garcia F., Chaty S., Chassande-Mottin E., Simaz Bunzel A., 2022, arXiv, arXiv:2206.03904 

\bibitem[Fryer et al.(2012)]{Fryeretal2012} Fryer, C.~L., Belczynski, K., Wiktorowicz, G., Dominik, M., Kalogera, V., \& Holz, D.~E.\ 2012, \apj, 749, 91

\bibitem[Gal-Yam(2012)]{GalYam2012} Gal-Yam, A.\ 2012, Science, 337, 927

\bibitem[Gilkis(2018)]{Gilkis2018} Gilkis, A.\ 2018, \mnras, 474, 2419. doi:10.1093/mnras/stx2934

\bibitem[Gilkis \& Soker(2015)]{GilkisSoker2015} Gilkis, A., \& Soker, N.\ 2015, \apj, 806, 28

\bibitem[\protect\citeauthoryear{Gilkis, Soker, \& Papish}{2016}]{Gilkisetal2016} Gilkis A., Soker N., Papish O., 2016, ApJ, 826, 178. doi:10.3847/0004-637X/826/2/178

\bibitem[\protect\citeauthoryear{Gofman \& Soker}{2019}]{GofmanSoker2019} Gofman R.~A., Soker N., 2019, MNRAS, 488, 5854. doi:10.1093/mnras/stz2179

\bibitem[\protect\citeauthoryear{Gomez et al.}{2022}]{Gomezetal2022} Gomez S., Berger E., Nicholl M., Blanchard P.~K., Hosseinzadeh G., 2022, arXiv, arXiv:2204.08486
 
\bibitem[Greiner et al.(2015)]{Greineretal2015} Greiner, J., Mazzali, P.~A., Kann, D.~A., et al.\ 2015, \nat, 523, 189

\bibitem[\protect\citeauthoryear{Grichener, Cohen, \& Soker}{2021}]{Gricheneretal2021} Grichener A., Cohen C., Soker N., 2021, ApJ, 922, 61. doi:10.3847/1538-4357/ac23dd

\bibitem[Grichener \& Soker(2017)]{GrichenerSoker2017} Grichener, A., \& Soker, N.\ 2017, \mnras, 468, 1226

\bibitem[Hillel et al.(2022)]{Hilleletal2022} Hillel, S., Schreier, R., \& Soker, N.\ 2022, \mnras, 514, 3212. doi:10.1093/mnras/stac1341

\bibitem[\protect\citeauthoryear{Hirai, Podsiadlowski, \& Yamada}{2018}]{Hiraietal2018} Hirai R., Podsiadlowski P., Yamada S., 2018, ApJ, 864, 119. doi:10.3847/1538-4357/aad6a0

\bibitem[\protect\citeauthoryear{Hirai \& Yamada}{2015}]{HiraiYamada2015} Hirai R., Yamada S., 2015, ApJ, 805, 170. doi:10.1088/0004-637X/805/2/170

\bibitem[Hosseinzadeh et al.(2022)]{Hosseinzadehetal2021} Hosseinzadeh, G., Berger, E., Metzger, B.~D., Gomez S., Nicholl M., Blanchard P.,\ 2022, \apj, 933, 14. doi:10.3847/1538-4357/ac67dd

\bibitem[Houck \& Chevalier(1991)]{HouckChevalier1991} Houck, J.~C., \& Chevalier, R.~A.\ 1991, \apj, 376, 234

\bibitem[Inserra et al.(2016)]{Inserraetal2016}  Inserra, C., Bulla, M., Sim, S.~A., \& Smartt, S.~J.\ 2016, \apj, 831, 79


\bibitem[Janka et al.(2017)]{Jankaetal2017} Janka, H.-T., Gabler, M., \& Wongwathanarat, A.\ 2017, Supernova 1987A:30 years later - Cosmic Rays and Nuclei from Supernovae and their Aftermaths, 331, 148

\bibitem[Kangas et al.(2017)]{Kangasetal2017} Kangas, T., Blagorodnova, N., Mattila, S., et al.\ 2017, \mnras, 469, 1246

\bibitem[\protect\citeauthoryear{Kaplan \& Soker}{2020a}]{KaplanSoker2020} Kaplan N., Soker N., 2020a, MNRAS, 492, 3013. doi:10.1093/mnras/staa020

\bibitem[\protect\citeauthoryear{Kaplan \& Soker}{2020b}]{KaplanSoker2020Jetshaped} Kaplan N., Soker N., 2020b, MNRAS, 494, 5909. doi:10.1093/mnras/staa1201

\bibitem[Kasen et al.(2016)]{Kasenetal2016} Kasen, D., Metzger, B.~D., \& Bildsten, L.\ 2016, \apj, 821, 36

\bibitem[\protect\citeauthoryear{Kashi, Michaelis, \& Kaminsky}{2022}]{Kashietal2022} Kashi A., Michaelis A., Kaminetsky Y., 2022, arXiv:2207.01990

\bibitem[\protect\citeauthoryear{Larsson et al.}{2021}]{Larssonetal2021} Larsson J., Sollerman J., Lyman J.~D., Spyromilio J., Tenhu L., Fransson C., Lundqvist P., 2021, ApJ, 922, 265. doi:10.3847/1538-4357/ac2a41

\bibitem[\protect\citeauthoryear{Li et al.}{2020}]{Lietal2020} Li L., Dai Z.-G., Wang S.-Q., Zhong S.-Q., 2020, ApJ, 900, 121. doi:10.3847/1538-4357/aba95b

\bibitem[\protect\citeauthoryear{Lin et al.}{2020}]{Linetal2020} Lin W.~L., Wang X.~F., Li W.~X., Zhang J.~J., Mo J., Sai H.~N., Zhang X.~H., et al., 2020, MNRAS, 497, 318. doi:10.1093/mnras/staa1918

\bibitem[Liu et al.(2017)]{LiuModjaz2017} Liu, Y.-Q., Modjaz, M., \& Bianco, F.~B.\ 2017, \apj, 845, 85

\bibitem[\protect\citeauthoryear{Liu et al.}{2015}]{LiuZWetal2015} Liu Z.-W., Tauris T.~M., R{\"o}pke F.~K., Moriya T.~J., Kruckow M., Stancliffe R.~J., Izzard R.~G., 2015, A\&A, 584, A11. doi:10.1051/0004-6361/201526757

\bibitem[Livio et al.(1986)]{Livioetal1986} Livio, M., Soker, N., de Kool, M., \& Savonije, G.~J., \ 1986, \mnras, 222, 235. doi:10.1093/mnras/222.2.235

\bibitem[Lopez et al.(2014)]{Lopezetal2014} Lopez, L.~A., Castro, D., Slane, P.~O., Ramirez-Ruiz, E., \& Badenes, C.\ 2014, \apj, 788, 5

\bibitem[\protect\citeauthoryear{Lunnan et al.}{2020}]{Lunnanetal2020} Lunnan R., Yan L., Perley D.~A., Schulze S., Taggart K., Gal-Yam A., Fremling C., et al., 2020, ApJ, 901, 61. doi:10.3847/1538-4357/abaeec

\bibitem[MacLeod \& Ramirez-Ruiz(2015a)]{MacLeodRamirezRuiz2015a} MacLeod, M., \&  {Ramirez-Ruiz}, E.\ 2015a, \apjl, 798, L19

\bibitem[MacLeod \& Ramirez-Ruiz(2015b)]{MacLeodRamirezRuiz2015b} MacLeod, M., \& Ramirez-Ruiz, E.\ 2015b, \apj, 803, 41

\bibitem[\protect\citeauthoryear{Margutti et al.}{2018}]{Marguttietal2018} Margutti R., Chornock R., Metzger B.~D., Coppejans D.~L., Guidorzi C., Migliori G., Milisavljevic D., et al., 2018, ApJ, 864, 45. doi:10.3847/1538-4357/aad2df

\bibitem[\protect\citeauthoryear{Moriya, Marchant, \& Blinnikov}{2020}]{Moriyaetal2020} Moriya T.~J., Marchant P., Blinnikov S.~I., 2020, A\&A, 641, L10. doi:10.1051/0004-6361/202038903

\bibitem[Mauerhan et al.(2017)]{Mauerhanetal2017} Mauerhan, J.~C., Van Dyk, S.~D., Johansson, J., Hu, M., Fox, O.~D., Wang, L., Graham, M.~L., Filippenko, A.~V., \& Shivvers, I.\ 2017, \apj, 834, 118

\bibitem[Mazzali et al.(2017)]{Mazzalietal2017} Mazzali, P.~A., Sauer, D.~N., Pian, E., Deng, J., Prentice, S., Ben Ami, S., Taubenberger, S., \& Nomoto, K.\ 2017, \mnras, 469, 2498

\bibitem[Metzger et al.(2015)]{Metzgeretal2015} Metzger, B.~D., Margalit, B., Kasen, D., \& Quataert, E.\ 2015, \mnras, 454, 3311

\bibitem[Milisavljevic et al.(2013)]{Milisavljevic2013} Milisavljevic, D., Soderberg, A.~M., Margutti, R., et al.\ 2013, \apjl, 770, LL38

\bibitem[\protect\citeauthoryear{Nicholl}{2021}]{Nicholl2021} Nicholl M., 2021, A\&G, 62, 5.34. doi:10.1093/astrogeo/atab092

\bibitem[Nicholl et al.(2017)]{Nicholletal2017a} Nicholl, M., Berger, E., Margutti, R.,  Blanchard, P. K., Milisavljevic, D., Challis, P., Metzger, B. D., \& Chornock, R.\ 2017, \apjl, 835, L8

\bibitem[Nishimura et al.(2015)]{Nishimura2015} Nishimura, N., Takiwaki, T., \& Thielemann, F.-K.\ 2015, \apj, 810, 109

\bibitem[\protect\citeauthoryear{Ogata, Hirai, \& Hijikawa}{2021}]{Ogataetal2021} Ogata M., Hirai R., Hijikawa K., 2021, MNRAS, 505, 2485. doi:10.1093/mnras/stab1439

\bibitem[Orlando et al.(2021)]{Orlandoetal2021} Orlando, S., Wongwathanarat, A., Janka, H.-T., Miceli, M., Ono, M., Nagataki, S., Bocchino, F., \& Peres, G.\ 2021, \aap, 645, A66. doi:10.1051/0004-6361/202039335

\bibitem[Papish \& Soker(2011)]{PapishSoker2011} Papish, O., \& Soker, N.\ 2011, \mnras, 416, 1697

\bibitem[Papish et al.(2015)]{Papishetal2015} Papish, O., Soker, N., \& Bukay, I.\ 2015, \mnras, 449, 288

\bibitem[Paxton et al.(2011)]{Paxtonetal2011} Paxton, B., Bildsten, L., Dotter, A., et al.\ 2011, \apjs, 192, 3

\bibitem[Paxton et al.(2013)] {Paxtonetal2013} Paxton, B., Cantiello, M., Arras, P., et al. 2013, \apjs, 208, 4

\bibitem[Paxton et al.(2015)]{Paxtonetal2015} Paxton, B., Marchant, P., Schwab, J., et al.\ 2015, \apjs, 220, 15

\bibitem[Paxton et al.(2018)]{Paxtonetal2018} Paxton, B., Schwab, J., Bauer, E.~B., et al.\ 2018, \apjs, 234, 34

\bibitem[Paxton et al.(2019)]{Paxtonetal2019} Paxton, B., Smolec, R., Schwab, J., et al.\ 2019, \apjs, 243, 10, doi:10.3847/1538-4365/ab2241

\bibitem[\protect\citeauthoryear{Pellegrino et al.}{2022}]{Pellegrinoetal2022} Pellegrino C., Howell D.~A., Terreran G., Arcavi I., Bostroem K.~A., Brown P.~J., Burke J., et al., 2022, arXiv, arXiv:2205.07894

\bibitem[Prentice et al.(2021)]{Prenticeetal2021} Prentice, S.~J., Inserra, C., Schulze, S., Nicholl M., Mazzali P.~A., Vergani S.~D., Galbany L., et al.,\ 2021, \mnras, 508, 4342. doi:10.1093/mnras/stab2864

\bibitem[Reilly et al.(2017)]{Reilletal2017} Reilly, E., Maund, J.~R., Baade, D., Wheeler, J.~C., H\"oflich, P., Spyromilio, J., Patat, F., \& Wang, L.\ 2017, \mnras, 470, 1491  

\bibitem[Ricker \& Taam(2008)]{RickerTaam2008}  Ricker, P.~M., \& Taam, R.~E.\ 2008, \apjl, 672, L41

\bibitem[\protect\citeauthoryear{Rimoldi, Portegies Zwart, \& Rossi}{2016}]{Rimoldietal2016} Rimoldi A., Portegies Zwart S., Rossi E.~M., 2016, ComAC, 3, 2. doi:10.1186/s40668-016-0015-4
See: zRimoldi2016.pdf 

\bibitem[\protect\citeauthoryear{Sandoval et al.}{2021}]{Sandovaletal2021} Sandoval M.~A., Hix W.~R., Messer O.~E.~B., Lentz E.~J., Harris J.~A., 2021, ApJ, 921, 113. doi:10.3847/1538-4357/ac1d49

\bibitem[\protect\citeauthoryear{Schreier et al.}{2021}]{Schreieretal2021}  Schreier R., Hillel S., Shiber S., Soker N., 2021, MNRAS, 508, 2386. doi:10.1093/mnras/stab2687 

\bibitem[\protect\citeauthoryear{Shiber, Kashi, \& Soker}{2017}]{Shiberetal2017}  Shiber S., Kashi A., Soker N., 2017, MNRAS, 465, L54. doi:10.1093/mnrasl/slw208      
\bibitem[\protect\citeauthoryear{Shishkin \& Soker}{2021}]{ShishkinSoker2021} Shishkin D., Soker N., 2021, MNRAS, 508, L43. doi:10.1093/mnrasl/slab105

\bibitem[Shishkin \& Soker(2022)]{ShishkinSoker2022} Shishkin, D. \& Soker, N.\ 2022, \mnras. doi:10.1093/mnras/stac1075

\bibitem[Sobacchi et al.(2017)]{Sobacchietal2017} Sobacchi, E., Granot, J., Bromberg, O., \& Sormani, M.~C.\ 2017, \mnras, 472, 616

\bibitem[Soker(2016a)]{Soker2016Mag} Soker, N.\ 2016a, \na, 47, 88

\bibitem[\protect\citeauthoryear{Soker}{2016b}]{Soker2016Rev} Soker N., 2016b, NewAR, 75, 1. doi:10.1016/j.newar.2016.08.002

\bibitem[Soker(2017a)]{Soker2017Mag2} Soker, N.\ 2017a, \apjl, 839, L6

\bibitem[Soker(2017b)]{Soker2017RAA} Soker, N.\ 2017b, Research in Astronomy and Astrophysics (arXiv:1702.03451)

\bibitem[\protect\citeauthoryear{Soker}{2020}]{Soker2020} Soker N., 2020, ApJ, 902, 130. doi:10.3847/1538-4357/abb809

\bibitem[\protect\citeauthoryear{Soker}{2022a}]{Soker2022} Soker N., 2022a, RAA, 22, 035019. doi:10.1088/1674-4527/ac49e6

\bibitem[\protect\citeauthoryear{Soker}{2022b}]{Soker2022Boosting}  Soker N., 2022b, arXiv, arXiv:2202.05556 


\bibitem[\protect\citeauthoryear{Soker}{2022c}]{Soker2022LSNe} Soker N., 2022c, arXiv, arXiv:2205.09560 

\bibitem[\protect\citeauthoryear{Soker}{2022d}]{Soker2022bump} Soker N., 2022d, arXiv, arXiv:2207.02753 


\bibitem[\protect\citeauthoryear{Soker \& Gilkis}{2017}]{SokerGilkis2017} Soker N., Gilkis A., 2017, ApJ, 851, 95. doi:10.3847/1538-4357/aa9c83

\bibitem[Soker \& Gilkis(2018)]{SokerGilkis2018} Soker, N., \& Gilkis, A.\ 2018, \mnras, 475, 1198

\bibitem[Sollerman et al.(2019)]{Sollermanetal2019} Sollerman, J., Taddia, F., Arcavi, I., et al.\ 2019, \aap, 621, A30

\bibitem[Sollerman et al.(2022)]{Sollermanetal2022} Sollerman, J., Yang, S., Perley, D., Schulze S., Fremling C., Kasliwal M., Shin K., et al.,\ 2022, \aap, 657, A64. doi:10.1051/0004-6361/202142049

\bibitem[Sorokina et al.(2016)]{Sorokinaetal2016}  Sorokina, E., Blinnikov, S., Nomoto, K., Quimby, R., \& Tolstov, A.\ 2016, \apj, 829, 17

\bibitem[Sukhbold \& Woosley(2016)]{SukhboldWoosley2016} Sukhbold, T., \& Woosley, S.~E.\ 2016, \apjl, 820, L38

\bibitem[\protect\citeauthoryear{Urvachev et al.}{2021}]{Urvachevetal2021} Urvachev E., Shidlovski D., Tominaga N., Glazyrin S., Blinnikov S., 2021, ApJS, 256, 8. doi:10.3847/1538-4365/ac0972


\bibitem[\protect\citeauthoryear{van den Heuvel}{2019}]{vandenHeuvel2019} van den Heuvel E.~P.~J., 2019, IAUS, 346, 1. doi:10.1017/S1743921319001315

\bibitem[\protect\citeauthoryear{Villar, Nicholl, \& Berger}{2018}]{Villaretal2018} Villar V.~A., Nicholl M., Berger E., 2018, ApJ, 869, 166. doi:10.3847/1538-4357/aaee6a

\bibitem[Wang et al.(2016)]{Wangetal2016} Wang, S.~Q., Liu, L.~D., Dai, Z.~G., Wang, L.~J., \& Wu, X.~F.\ 2016, \apj, 828, 87

\bibitem[Wang et al.(2019)]{Wangetal2019RAA} Wang, S.-Q., Wang, L.-J., \& Dai, Z.-G.\ 2019, Research in Astronomy and Astrophysics, 19, 063. doi:10.1088/1674-4527/19/5/63

\bibitem[Wheeler et al.(2002)]{Wheeleretal2002}  Wheeler, J.~C., Meier, D.~L., \& Wilson, J.~R.\ 2002, \apj, 568, 807 

\bibitem[\protect\citeauthoryear{Wiktorowicz et al.}{2017}]{Wiktorowiczetal2017} Wiktorowicz G., Sobolewska M., Lasota J.-P., Belczynski K., 2017, ApJ, 846, 17. doi:10.3847/1538-4357/aa821d


\bibitem[Woosley(2018)]{Woosley2018} Woosley, S.~E.\ 2018, \apj, 863, 105.

\bibitem[Yu et al.(2017)]{Yuetal2017} Yu, Y.-W., Zhu, J.-P., Li, S.-Z., L{\"u}, H.-J., \& Zou, Y.-C.\ 2017, \apj, 840, 12  

\bibitem[\protect\citeauthoryear{Zuo \& Li}{2014}]{ZuoLi2014} Zuo Z.-Y., Li X.-D., 2014, ApJ, 797, 45. doi:10.1088/0004-637X/797/1/45


\end{thebibliography}
\end{document}